\documentclass[12pt]{iopart}
\usepackage{graphicx}
\usepackage{iopams}
\usepackage{amssymb}
\usepackage{multirow}

\newcommand{\ped}[1]{\ensuremath{_{\rm #1}}}
\newcommand{\apex}[1]{\ensuremath{^{\rm #1}}}
\begin{document}
\title{Theoretical explanation of electric field-induced superconductive critical temperature shifts in Indium thin films}
\author{G.A.~Ummarino}
\ead{giovanni.ummarino@polito.it}
\address{Istituto di Ingegneria e Fisica dei Materiali,
Dipartimento di Scienza Applicata e Tecnologia, Politecnico di
Torino, Corso Duca degli Abruzzi 24, 10129 Torino, Italy; National Research Nuclear University MEPhI (Moscow Engineering Physics Institute),
Kashira Hwy 31, Moskva 115409, Russia}
\author{D. Romanin}
\address{Istituto di Ingegneria e Fisica dei Materiali,
Dipartimento di Scienza Applicata e Tecnologia, Politecnico di
Torino, Corso Duca degli Abruzzi 24, 10129 Torino, Italy}
\begin{abstract}
We calculate the effect of a static electric field on the superconductive critical temperature of Indium thin films in the framework of proximity effect Eliashberg theory, in order to explain 60 years old experimental data.
Since in the theoretical model we employ all quantities of interest can be computed ab-initio (i.e. electronic densities of states, Fermi energy shifts and Eliashberg spectral functions), the only free parameter is in general the thickness of the surface layer where the electric field acts. However, in the weak electrostatic field limit Thomas-Fermi approximation is still valid and therefore no free parameters are left, as this perturbed layer is known to have a thickness of the order of the Thomas-Fermi screening length. We show that the theoretical model can reproduce experimental data, even when the magnitude of the induced charge densities are so small to be usually neglected.
\end{abstract}
%
\maketitle
\section{INTRODUCTION}
\label{intro}

In recent years, electrostatic fields have emerged as a powerful tool to control the physical properties of many different superconductive materials in the field-effect transistor (FET) architecture. Significant successes have been obtained by means of the field-induced ultrahigh surface charge doping attainable via the ionic gating technique, which allowed to efficiently tune the properties of different classes of superconductors. These included metallic superconductors \cite{ChoiAPL2014, PiattiJSNM2016, PiattiNbN}, transition-metal dichalcogenides \cite{LiNature2016, XiPRL2016, YoshidaAPL2016, LiNanoLett2019}, iron-based superconductors \cite{ShiogaiNatPhys2016, LeiPRL2016, HanzawaPNAS2016, ZhuPRB2017, MiyakawaPRM2018, KounoScirep2018, PiattiPRM2019} and cuprates \cite{cuprates1, cuprates2,cuprates3,cuprates4,cuprates5,cuprates6,cuprates7,cuprates8,cuprates9}. More recently, even conventional solid gating was shown to be suprisingly effective in tuning the superconductive properties of nanostructures of metallic superconductors \cite{DeSimoniNatNano2018, PaolucciNL2018, PaolucciPRAp2019}. However, the investigation of the effects of an electrostatic field on a superconductor dates back almost seventy years: in 1960, R.E. Glover, III and M. D. Sherrill\cite{glover} examined the effect of charging metals with the aid of a static electric field. They shoed that the conductivity of metals was modified and, for what concerns the superconductive materials, the transition temperature could be positively or negatively shifted. Measurements made in the superconductive transition region for five indium films showed in every case an increase in resistance with negative charging (i.e. by adding electrons) and a corresponding decrease with positive charging (i.e. by removing electrons). The measured resistance changes correspond to a decrease in transition temperature with negative charging. Thicknesses ranged from 60 to 120 \AA. A field of $2.6\times10^7$ V/m (approximately $3\times10^{-5}$ electron per atom) produced a shift in transition temperature on the order of $10^{-4}$ K. Up to now, as far as we know, a plausible explanation has not been given to these experimental results, and indeed the very possibility to describe the effects of an electrostatic field on a metallic superconductor in terms of charge doping has been recently called into question\cite{DeSimoniNatNano2018, PaolucciNL2018, PaolucciPRAp2019}.

In this work, we show instead a theoretical treatment to properly describe the effect of an electrostatic field on the superconductive properties of more complex materials developed in the framework of Eliashberg theory and successfully applied to Pb and MgB$_2$ \cite{ummarinoPb, dodoMgB2}. The further development of such a theoretical framework and its validation on different classes of superconductive materials is important in order to both quantitatively describe future experimental results and suggest a priori experimental conditions (e.g., number of carriers to induce, device thickness, etc.) for an optimal modulation of superconductive properties in the field-effect architecture \cite{dodoDiamond}.

The influence of an electrostatic field on a superconductive thin film with a thickness exceeding the electrostatic screening length can be modeled as follows \cite{ummarinoPb, dodoMgB2}. In the absence of an applied electric field the material is homogeneous and unperturbed. However when we turn on the electrostatic perturbation, the electric field penetrates in a surface layer of the thin film and thus identifies two spatial regions which form a junction between a superconductor and a normal metal in the temperature range $T\in[T_{c,s};T_{c,b}]$ ($T_{c,s}\ne T_{c,b}$, $s$ and $b$ indicate "surface" and "bulk" respectively). We can identify a perturbed surface layer ($T_c=T_{c,s}$), where the electric field modulates the carrier density (i.e. where the electric field induces a doping level per unitary cell $x$), and an underlying unperturbed bulk ($T_c=T_{c,b}$). In general, if we have a superconductor/normal metal junction, the proximity effect is observed as the opening of a finite superconductive gap in the normal metal together with its reduction in a thin region of the superconductor close to the junction. In the case of Indium, all input parameters of the theory are well known in literature \cite{glover,carbibastardo}.

The paper is organized as follow. In Sec.~\ref{sec:model} we show the model we use for the computation of the superconductive critical temperature, i.e. the one band s-wave Eliashberg equations with proximity effect. After that, in Sec.~\ref{sec:DFT} we expose the computational details used for ab-initio calculations. In Sec.~\ref{sec:results} we discuss our results on Indium thin films. Finally, conclusions are given in Sec.~\ref{sec:conclusions}.

\section{MODEL: PROXIMITY ELIASHBERG EQUATIONS}
\label{sec:model}
The model we employ calculates the critical temperature of the system by solving the one band s-wave Eliashberg equations  \cite{carbibastardo,ummarinorev} with proximity effect. In this case four coupled equations for the renormalization functions $Z_{s,b}(i\omega_{n})$ and gaps $\Delta_{s,b}(i\omega_{n})$ have to be solved ($\omega_{n}$ denotes the Matsubara frequencies). The set of equations with proximity effect on the imaginary-axis \cite{Mc,Carbi1,Carbi2,Carbi3,kresin} is:
\begin{eqnarray}
&&\omega_{n}Z_{b}(i\omega_{n})=\omega_{n}+ \pi T\sum_{m}\Lambda^{Z}_{b}(i\omega_{n},i\omega_{m})N^{Z}_{b}(i\omega_{m})+\nonumber\\
&&+\Gamma\ped{b} N^{Z}_{s}(i\omega_{n})
\label{eq:EE1}
\end{eqnarray}
\begin{eqnarray}
&&Z_{b}(i\omega_{n})\Delta_{b}(i\omega_{n})=\pi
T\sum_{m}\big[\Lambda^{\Delta}_{b}(i\omega_{n},i\omega_{m})-\mu^{*}_{b}(\omega_{c})\big]\times\nonumber\\
&&\times\Theta(\omega_{c}-|\omega_{m}|)N^{\Delta}_{b}(i\omega_{m})
+\Gamma\ped{b} N^{\Delta}_{s}(i\omega_{n})\phantom{aaaaaa}
 \label{eq:EE2}
\end{eqnarray}
\begin{eqnarray}
&&\omega_{n}Z_{s}(i\omega_{n})=\omega_{n}+ \pi T\sum_{m}\Lambda^{Z}_{s}(i\omega_{n},i\omega_{m})N^{Z}_{s}(i\omega_{m})+\nonumber\\
&&\Gamma\ped{s} N^{Z}_{b}(i\omega_{n})
\label{eq:EE3}
\end{eqnarray}
\begin{eqnarray}
&&Z_{s}(i\omega_{n})\Delta_{s}(i\omega_{n})=\pi
T\sum_{m}\big[\Lambda^{\Delta}_{s}(i\omega_{n},i\omega_{m})-\mu^{*}_{s}(\omega_{c})\big]\times\nonumber\\
&&\times\Theta(\omega_{c}-|\omega_{m}|)N^{\Delta}_{s}(i\omega_{m})
+\Gamma\ped{s}N^{\Delta}_{b}(i\omega_{n})\phantom{aaaaaa}
 \label{eq:EE4}
\end{eqnarray}

where $\omega_{c}$ is a cutoff energy at least three times larger than the maximum phonon energy,  $\mu^{*}_{s(b)}$ are the Coulomb pseudopotentials in the surface and in the bulk respectively and $\Theta$ is the Heaviside function.
Moreover:

\begin{equation}
N^{\Delta}_{s(b)}(i\omega_{m})=\Delta_{s(b)}(i\omega_{m})/
{\sqrt{\omega^{2}_{m}+\Delta^{2}_{s(b)}(i\omega_{m})}}
\end{equation}
\begin{equation}
N^{Z}_{s(b)}(i\omega_{m})=\omega_{m}/{\sqrt{\omega^{2}_{m}+\Delta^{2}_{s(b)}(i\omega_{m})}}
\end{equation}
\begin{equation}
\Gamma_{s(b)}=\pi|t|^{2}Ad_{b(s)}N_{b(s)}(0)
\label{eq:EE6}
\end{equation}

with the relation $\frac{\Gamma_{s}}{\Gamma_{b}}=\frac{d_{b}N_{b}(0)}{d_{s}N_{s}(0)}$, where $A$ is the junction cross-sectional area, $d_{s}$ and $d_{b}$ are the surface and bulk layer thicknesses respectively, such that ($d_{s}+d_{b}=d$ where $d$ is the total film thickness) and $N_{s(b)}(0)$ are the densities of states at the Fermi level $E_{F,s(b)}$ for the surface and bulk material.
Finally:
\begin{equation}
\Lambda_{s(b)}(i\omega_{n},i\omega_{m})=2
\int_{0}^{+\infty}d\Omega \Omega
\alpha^{2}_{s(b)}F(\Omega)/[(\omega_{n}-\omega_{m})^{2}+\Omega^{2}]
\end{equation}

where $\alpha^{2}_{s(b)}F(\Omega)$ are the electron-phonon spectral functions.

We expect to have a nearly ideal interface between the surface and bulk layers since we only consider electrostatic perturbations to the system, therefore we assume the transmission matrix $|t|^{2}=1$. This assumption is supported by experimental findings on niobium nitride \cite{PiattiNbN}, where the experimental doping dependence of $T_c$ turned out to be compatible with a high interface transparency.\\

The electron-phonon coupling constants are defined as
\begin{equation}
\lambda_{s(b)}=2\int_{0}^{+\infty}d\Omega\frac{\alpha^{2}_{s(b)}F(\Omega)}{\Omega}\nonumber\\
\end{equation}
and the representative energies as
\begin{equation}
\omega_{ln,s(b)}=\exp\Bigl\{\frac{2}{\lambda_{s(b)}}\int_{0}^{+\infty}d\Omega ln\Omega \frac{\alpha^{2}_{s(b)}F(\Omega)}{\Omega}\nonumber\Bigr\}\\
\end{equation}

In order to solve this set of coupled equations, eleven input parameters are needed: the two electron-phonon spectral fuctions $\alpha^{2}_{s(b)}F(\Omega)$, the two Coulomb pseudopotentials $\mu^{*}_{s(b)}$, the values of the normal density of states at the Fermi level $N_{s(b)}(0)$, the shift of the Fermi energy $\Delta E_{F}=E_{F,s}-E_{F,b}$ that enters in the calculation of the surface Coulomb pseudopotential \cite{ummarinoPb}, the thickness of the surface layer $d_{s}$, the total film thickness $d$ and the junction cross-sectional area $A$. The values of $d$ and $A$ are experimental data. The exact value of $d_{s}$ is in general difficult to be determined \emph{a priori} in the case of metals, in particular for very strong applied electric fields. However, for low magnitude values of the perturbation $d_{s}$ can be taken to be equal to the Thomas-Fermi screening length $d_{TF}$\cite{piattidodoNbN}.

Moreover we point out that usually, when the shift of chemical potential ($\Delta E_{F}$) is very small in comparison with the Fermi energy ($8630$ meV), the effect of electrostatic doping on $\mu^{*}$ can be neglected \cite{ummarinoPb} and, as a consequence, $\mu^{*}_{b}=\mu^{*}_{s}$.

\section{AB-INITIO CALCULATION OF $\alpha^{2}_{s(b)}F(\Omega)$, $\Delta E_{F}$ and $N_{s(b)}(0)$}
\label{sec:DFT}
\begin{figure}
\begin{center}
\includegraphics[keepaspectratio, width=0.4\columnwidth]{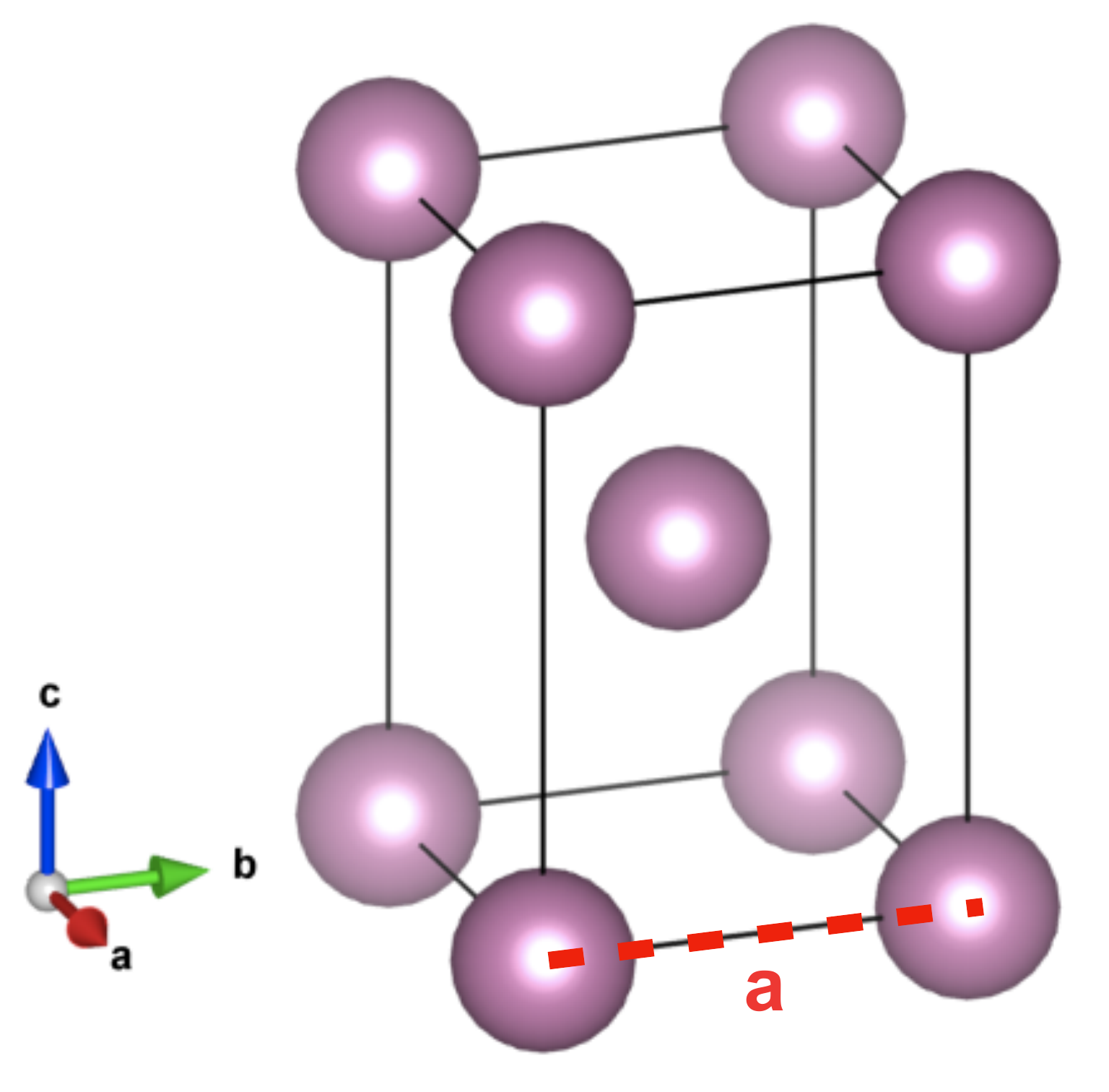}
\vspace{-5mm} \caption{(Color online)
 Atomic structure of body-centered tetragonal Indium. Red dashed lines denotes the lattice parameter.
 }\label{Figure0}
\end{center}
\end{figure}
We model our system as a junction between a perturbed surface layer and an unperturbed underlying bulk. Electronic and vibrational properties for both sub-systems are computed for bulk Indium in its body-centered tetragonal structure, which has one atom per unit cell (see Fig.~\ref{Figure0}. For the affected surface layer, we simulate the additional charge using a jellium model (i.e. with a uniform distribution of charges compensated by a background of opposite sign). The induced charge in the material is treated assuming that the additional carriers spread uniformly inside the surface layer of thickness $d_s$, which is justified for not too high values of the applied electric field: indeed, in this case Thomas-Fermi approximation is still valid and the thickness of the perturbed surface layer is taken to be of the order of Thomas-Fermi screening length\cite{dodoMgB2,piattidodoNbN}. As a matter of fact, in a field-effect device, it is possible to induce carrier density \emph{per unit surface}, $n_{2D}$ by tuning the polarization of the gate electrode. Within our approximations, the induced carrier density \emph{per unit volume} can be obtained as $x =\Delta n_{2D}/d_s$. However if we were to investigate higher applied electric field (and, consequently, higher values of the induced charge densities) a suitable model of field-effect geometry should be applied\cite{brumme,thibault}.

The ab-initio computations of relevant quantities is performed in the framework of plane-wave pseudopotential density functional theory (DFT) as implemented in the Quantum ESPRESSO \cite{QE1,QE2}. In order to model the exchange-correlation interaction in Indium we exploited a Perdew-Burke-Ernxerhof (PBE) generalized gradient approximation. The effective interaction between core and valence electrons is treated with a scalar relativistic ultrasoft pseudopotential.

Wave functions of valence electrons are expanded in a plane-wave basis up to an energy cutoff of 48 Ry, while the density cutoff is set to 190 Ry. In order to reproduce already existing computations for unperturbed bulk Indium \cite{rudin}, the Brillouin zone is sampled on a uniform grid of $16\times16\times16$ k-points (using the Monkhorst-Pack grid algorithm), with a Methfessel-Paxton smearing \cite{mp} of 0.04 Ry. In order to compute densities of states at the Fermi level we use instead a uniform grid of $32\times32\times32$ k-points with tetrahedra smearing (which assures better convergence of results). Convergence of the self-consistent solution of Kohn-Sham equations is set to $10^{-9}$ Ry on the total energy. The same parameters are then used also for the doped bulk Indium, which was not yet investigated in Ref.~\cite{rudin}, upon checking convergence on total energy per atom ($E_{tot}/$atom $\le1$ mRy). The lattice parameter of undoped bulk Indium is computed by letting the structure relax towards zero interatomic forces, obtaining $a_{th}=3.30$ \\A that is $1.5\%$ larger than the experimental value $a_{exp}=3.25$\cite{expIn}.

Vibrational properties are then computed in the framework of density functional perturbation theory (DFPT) on a uniform grid of $8\times8\times8$ q-points. Electron-phonon matrix elements $g_{\bf{k},\bf{k}+\bf{q}}^{\nu}$ are computed on the coarse q-grid for each phonon mode $\nu$ and interpolated over the Brillouin zone as in Ref.~\cite{phononint}. Eliashberg spectral functions are computed using:

\begin{equation}
\alpha^{2}_{s(b)}F(\Omega) = \frac{1}{N_{s(b)}(0)N_kN_q}\sum_{\bf{q}\nu}\delta(\hbar\omega-\hbar\omega_{\bf{q}\nu})\sum_{\bf{k}}\|g_{\bf{k},\bf{k}+\bf{q}}^{\nu}\|^2\delta(\epsilon_{\bf{k}})\delta(\epsilon_{\bf{k}+\bf{q}})
\end{equation}

where $N_{s(b)}$ is the total density of space per spin at the Fermi energy ($E_F=0$), $N_q$ and $N_k$ are the number of q- and k-points considered in the first Brillouin zone for the computations and the energy $\epsilon$ is measured from the Fermi level. The $\delta(\hbar\omega-\hbar\omega_{\bf{q}\nu})$ is approximated with a Gaussian smearing of $0.05$ Ry.

\section{RESULTS AND DISCUSSION}
\label{sec:results}
In Eliashberg theory the superconductive critical temperature is an increasing function of both $\omega_{ln}$ (i.e. the representative phonon energy) and the electron-phonon coupling constant $\lambda$. This is can be better seen in the semi-empirical Allen-Dynes formula\cite{AllenDynes}, which is a limit of Eliashberg theory:

\begin{equation}
T_c = \frac{\omega_{ln}}{1.2}exp\left(\frac{1.04(1+\lambda)}{\lambda-\mu^{*}(1+0.62\lambda)}\right)
\end{equation}
Thus the enhancement or suppression of $T_c$ depends on which of the two contributions is dominant: the optimal way for having the largest possible critical temperature in a field-effect doped material would be to have a strong increase of $\lambda$ and $\omega_{ln}$ concurrently. Moreover, we need to point out that while $\omega_{ln}$ and $\lambda$ can be computed ab-initio, the effective electron-electron interaction $\mu^{*}$ is a parameter which has to be tuned ad-hoc by suitable criteria.

\begin{figure}
\begin{center}
\includegraphics[keepaspectratio, width=\columnwidth]{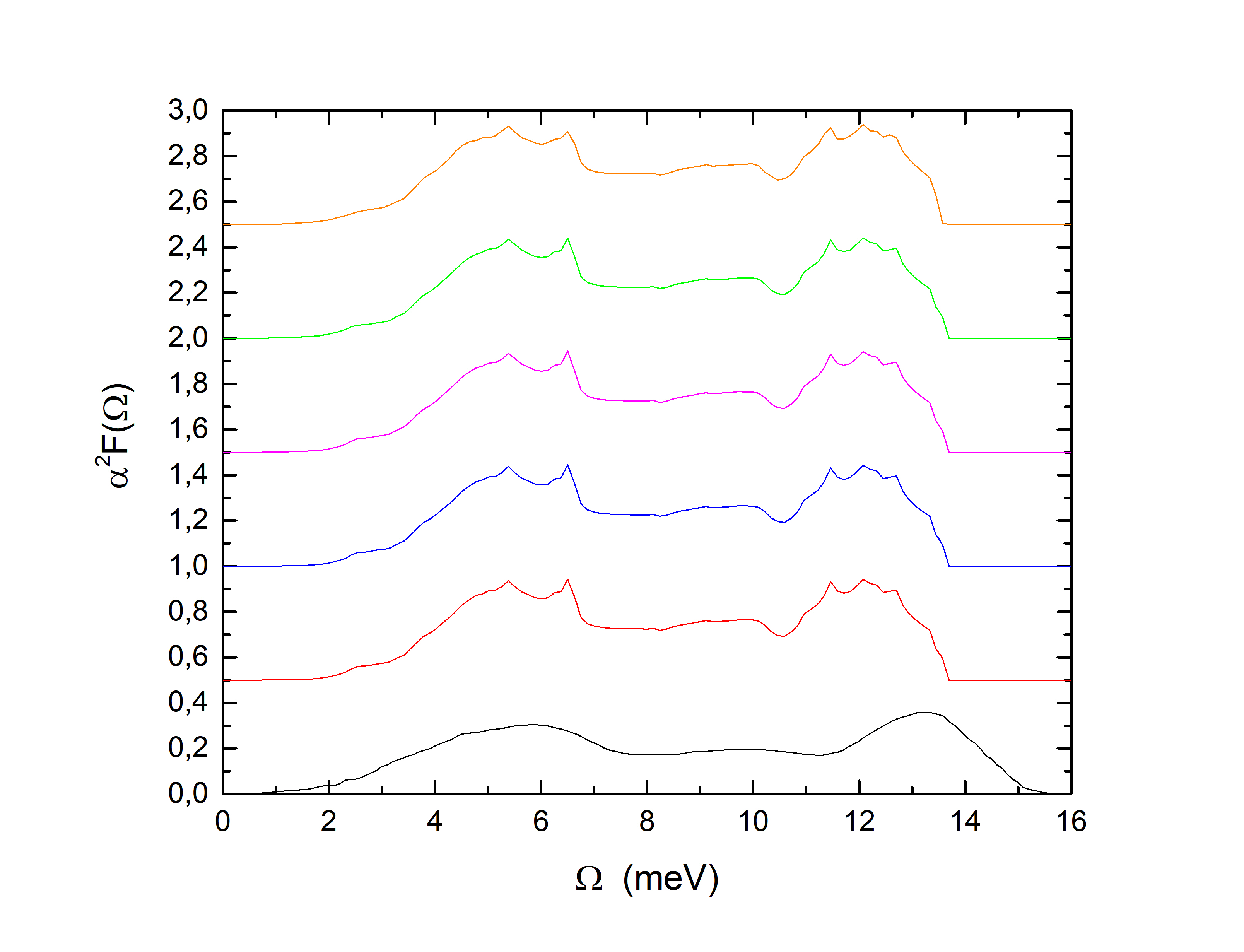}
\vspace{-5mm} \caption{(Color online)
 Calculated electron-phonon spectral function of Indium for five different values of charge doping (electrons/unitary cell): 0.00 (red solid line), $3\cdot 10^{-5}$ (blue solid line), $3\cdot 10^{-4}$ (violet solid line), $3\cdot 10^{-3}$ (green solid line) $3\cdot 10^{-2}$ (orange solid line). We also show the experimental electron-phonon spectral function determined via tunneling measurements \cite{rudin} (black solid line). All curves are shifted
 by a constant offset equal to 0.5.
 }\label{Figure1}
\end{center}
\end{figure}

In the case of Indium, we first compute the undoped $\alpha^{2}_{b}F(\Omega)$ that gives a corresponding electron-phonon coupling $\lambda_{b}=0.8728$. Then we solve one-band s-wave Eliashberg equations forcing $T_{c,b}$ to its experimental value \cite{carbibastardo} $T_{c,b} = 3.4$ K. Assuming a cutoff energy $\omega_{c} = 50$ meV and a maximum energy $\omega_{max} = 60$ meV in the Eliashberg equations, we are thus able to determine the bulk Coulomb pseudopotential to be $\mu^{*}_{b} = 0.170807$. After that, we move to study the effect of doping on electronic and vibrational properties of Indium. In Fig.~\ref{Figure1} we show the calculated electron-phonon spectral functions $\alpha^2F(\Omega)$ for increasing doping, i.e.  $x$ for $x=0.0, 0.00003, 0.0003, 0.003$  and $0.03$ e\apex{-}/atom while in Tab.~\ref{tab:tab1} we summarize all the input parameters for the proximity Eliashberg equations as obtained from DFT computations.\\

\begin{table*}
\begin{center}
\begin{tabular}{|c|c|c|c|c|c|}
  \hline
  $x (e^{-}/cell)$   & $\lambda$         & $\omega_{ln}$ $(meV)$  & $N(0)$ $states/eV/spin/cell$ & $\Delta E_{F}$ $(meV)$   & $T_{c}$  $(K)$ \\
  0.0           & 0.8728            & 6.4177               & 0.42390               &   0.0              & 3.40000 \\
  $3\times10^{-5}$           & 0.8724            & 6.4189               & 0.42410               & 1.0              & 3.39683 \\
  $3\times10^{-4}$           &  0.8730            & 6.4071             & 0.42390               & 2.0              & 3.39595 \\
  $3\times10^{-3}$           &  0.8745            & 6.3744             & 0.42310               & 4.0              & 3.39012 \\
  $3\times10^{-2}$           & 0.8768           & 6.3214              & 0.42060               & 43.2              & 3.38063 \\
  \hline
\end{tabular}
\caption{Input parameters calculated by DFT and $T_c$ calculated by Eliashberg theory without proximity effect.}\label{tab:exp}
\label{tab:tab1}
\end{center}
\end{table*}
\begin{figure}
\begin{center}
\includegraphics[keepaspectratio, width=\columnwidth]{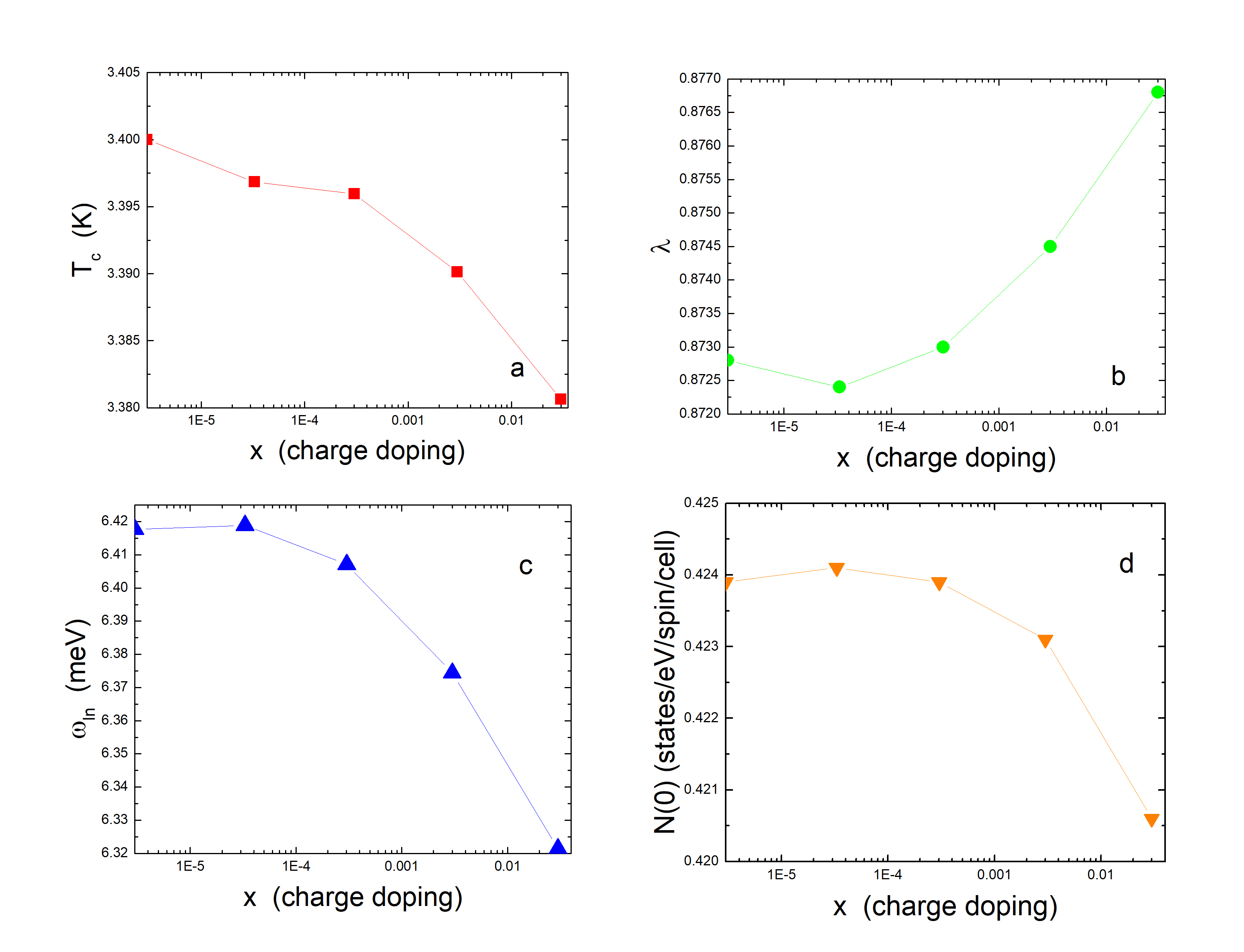}
\vspace{-5mm} \caption{(Color online)
 Paneal a: calculated critical temperature versus charge doping x (the scale is $x+3\cdot10^{-6}$ for graphic reasons) for a system without proximity effect; panel b: calculated electron-phonon coupling constant versus charge doping; panel c: calculated representative energy $\omega_{ln}$ versus charge doping; panel d: calculated normal density of states at the Fermi level versus charge doping. All lines act as guides to the eye. The graphs are in logarithmic scale.
 }\label{Figure2}
\end{center}
\end{figure}
With the first doping value we can observe a slight increase of $\omega_{ln}$ and a decrease of the electron-phonon coupling constant $\lambda$. However, moving to higher carrier densities, the value of $\omega_{ln}$ decreases while that of $\lambda$ increases. The density of states at the Fermi level has the same dependence on doping as $\lambda$ (see Fig.~\ref{Figure2} panel b,c and d). In the case of Indium, the contribution from the decreasing $\lambda$ is dominant over the increasing $\omega_{ln}$, giving rise to a net reduction of the superconductive critical temperature $T_{c,s}$ (as we report in Fig.~\ref{Figure2} panel a).

\begin{figure}[b]
\begin{center}
\includegraphics[keepaspectratio, width=0.9\columnwidth]{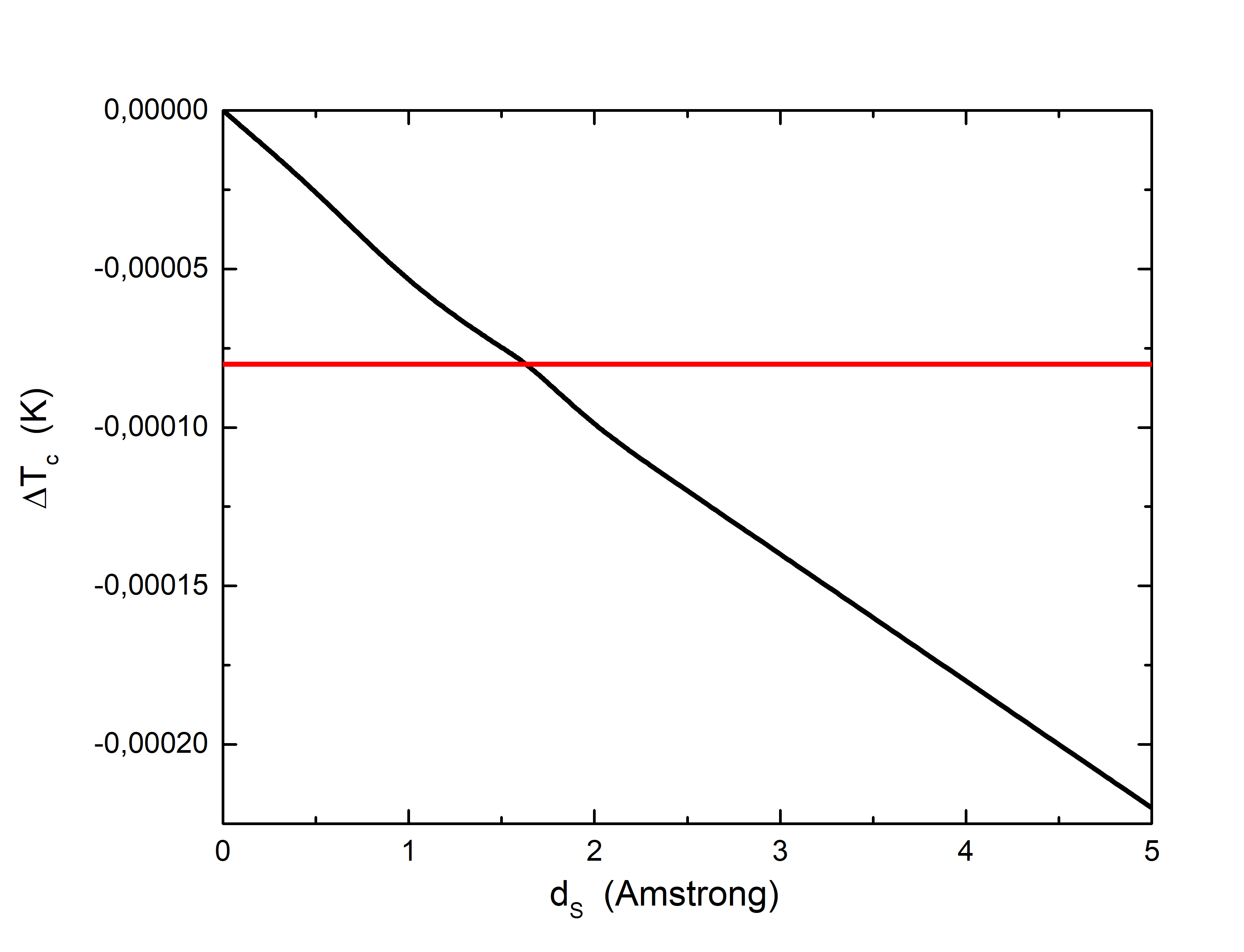}
\vspace{-5mm} \caption{(Color online)
 Calculated critical temperature versus surface layer thickness $d_{s}$ (black line) for a film of thickness $d=7$ nm and charge doping (electrons/unitary cell) $x=0.00003$. The red line is the experimental data. \cite{glover}.
 }\label{Figure3}
\end{center}
\end{figure}

We can now consider the behavior of the junction between the perturbed surface layer ($T_{c,s}$) and the unperturbed bulk ($T_{c,b}$). However, we first have to select a suitable value for $d_s$. Close to $T_c$, the screening is dominated by unpaired electrons since the superfluid density is small\cite{HirschPRB2004}. Thus, we can set $d_s=d_{TF}$, i.e. equal to Thomas-Fermi screening length, which as we already discussed is a good approximation for low electric fields that build up in the electric double layer \cite{PiattiNbN}. In the case of Indium, $d_{TF}=0.114$ nm, computed using relative dielectric constant \cite{epsilon} ($\varepsilon_{r}=9.3$), unit cell volume (27.2839 \AA$^{3}$) and density of states at the Fermi level (0.42390 $states/eV/spin/cell$). However, in order to exactly reproduce the superconductive critical temperature shift measured in Ref.~\cite{glover}, $d_s=0.165$ nm (see Fig.~\ref{Figure3}) that is nevertheless in agreement with the theoretical value of $d_{TF}$. This also justifies the Thomas-Fermi approximation for our calculations.

\begin{figure}
\begin{center}
\includegraphics[keepaspectratio, width=0.9\columnwidth]{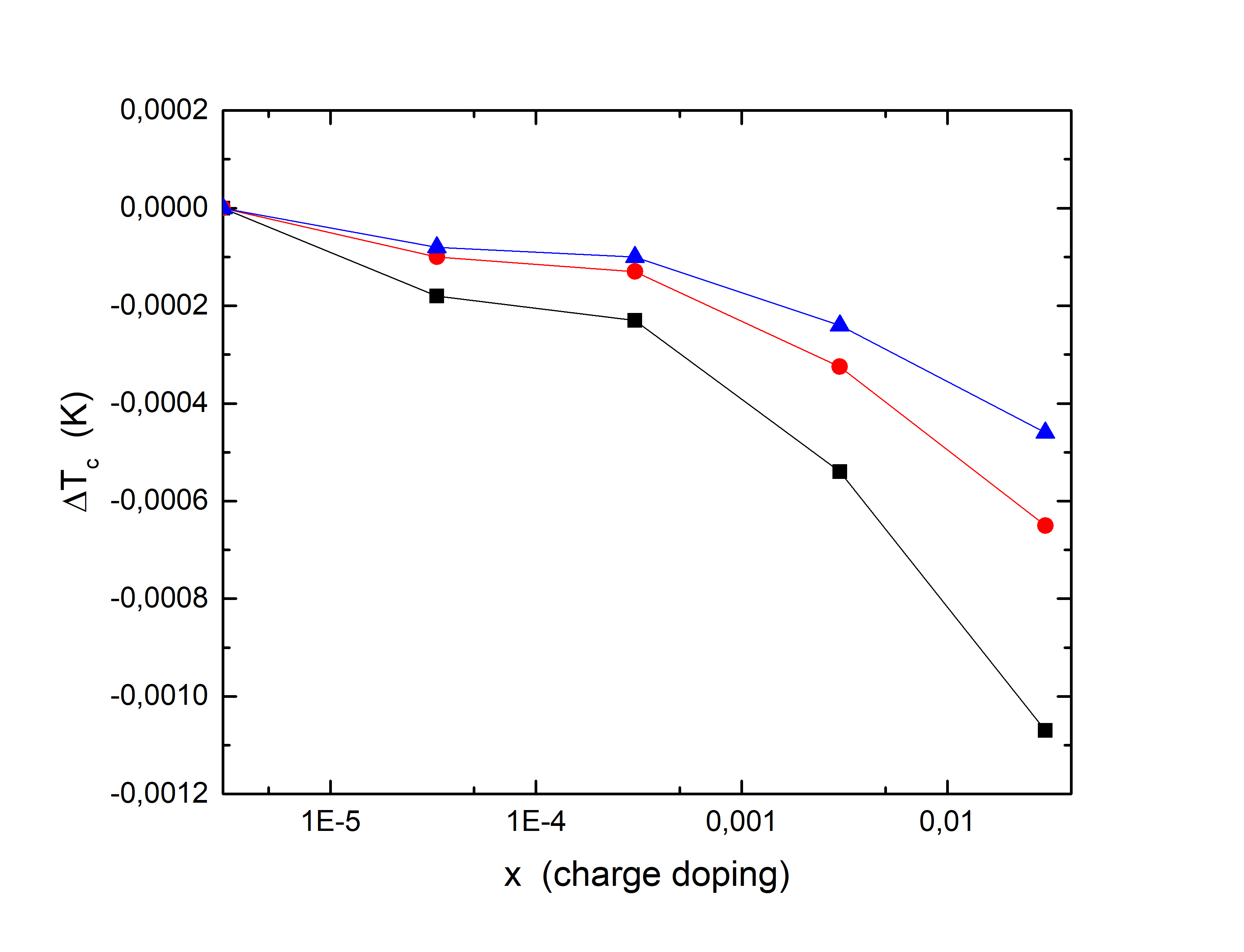}
\vspace{-5mm} \caption{(Color online)
 Calculated critical temperature versus charge doping x (the scale is $x+3\cdot10^{-6}$) with surface layer thickness $d_{s}=0.165$ nm for three different values of film thickness $d=3$ nm (black squares), $d=5$ nm (red circles), $d=7$ nm (blue up triangles). All lines act as guides to the eye. The scale is logarithmic.
 }\label{Figure4}
\end{center}
\end{figure}

In Fig. \ref{Figure4} we plot the results of proximity-coupled Eliashberg calculations for $\Delta T_{c}$ as a function of increasing electron doping and for three different film thicknesses $d = 3, 5, 7$ nm. For all three cases we assume that the Thomas-Fermi model still holds ($d_{s} = d_{TF}$) and that the junction area \cite{dodoMgB2} is $A = 10^{-6}$ m\apex{2}.

Qualitatively, in proximized films of any thickness,  $T_{c}$ decreases with increasing doping level with the same trend observed in the homogeneous case. However, the presence of a coupling between surface and bulk induced by the proximity effect gives rise to a striking difference with respect to the homogeneous case: $T_{c}$ strongly depends on film thickness in the doped films.
Indeed, the variations of critical temperature are greatly suppressed with increasing film thickness. We have not calculated the critical temperature for monolayer films since the approximations of the model would no longer apply in this case: in particular the unperturbed electron-phonon spectral function would be different from the bulk-like one we employed in our calculations \cite{Pratappone}.

\section{CONCLUSIONS}
\label{sec:conclusions}

In this work we have given a theoretical explanation to superconductive transition temperature shifts due to a static electric field measured in Indium thin films\cite{glover}. In order to do so we solved one band s-wave Eliashberg equations with proximity effect, whose input parameters were computed by means of density functional theory (DFT).

The system was modeled with a surface layer of thickness $d_s$ (whose charge density can be perturbed by a static electric field) and an underlying unperturbed bulk. Thanks to proximity effect these two subsystems are linked and affect each other in a non trivial way. Usually the thickness of the surface layer is difficult to determine a priori \cite{PiattiNbN} and a suitable model for field-effect geometry should be employed in DFT computations\cite{brumme,thibault}. However in the weak electric field limit, we can set $d_s$ equal to Thomas-Fermi screening length $d_{TF}$, thus getting rid of any free parameter \cite{dodoMgB2}.

The theoretical approach we emplyed, which has already given successful results in Pb\cite{ummarinoPb} and MgB$_2$\cite{dodoMgB2}, is able to reproduce experimental results of Ref.~\cite{glover}. As a consequence the Thomas-Fermi approximation is justified. Moreover we showed that the shifts of the superconductive critical temperature strongly depend on the total thickness of Indium thin films: indeed $\Delta T_c$ tends to be suppressed as $d$ is increased.

As a final remark, we have to stress the fact that in this work we show how induced charged densities which are usually considered negligible ($x=3\times10^{-5}$ electrons/atom) can have instead a relevant role on electronic and vibrational properties. This is a point which was recently questioned in Ref.~\cite{DeSimoniNatNano2018, PaolucciNL2018, PaolucciPRAp2019}.

\ackn
G.A.U. acknowledges support from the MEPhI Academic Excellence Project (Contract No. 02.a03.21.0005). Computational resources were provided by hpc@polito, which is a project of Academic Computing within the Department of Control and Computer Engineering at the Politecnico di Torino (http://hpc.polito.it). The authors thank E. Piatti for the suggestions.\\


\end{document}